\documentstyle[emulateapj,apjfonts,psfig,graphicx,epsfig]{article}

\def\sn{SN\,{2002ap}}
\def\ra#1#2#3{#1$^{\rm h}$#2$^{\rm m}$#3$^{\rm s}$}
\def\dec#1#2#3{$#1^\circ#2'#3''$}
\def\ale{\mathrel{\hbox{\rlap{\hbox{\lower4pt\hbox{$\sim$}}}\hbox{$<$}}}}
\def\age{\mathrel{\hbox{\rlap{\hbox{\lower4pt\hbox{$\sim$}}}\hbox{$>$}}}}

\begin{document}
 
\title{\large The Radio Evolution of the Ordinary Type Ic SN\,2002ap}

\author{E. Berger\altaffilmark{1} and S. R. Kulkarni\altaffilmark{1}}
\affil{Division of Physics, Mathematics, and Astronomy,
California Institute of Technology 105-24, Pasadena, CA 91125}
\author{R. A. Chevalier\altaffilmark{2}}
\affil{Department of Astronomy, University of Virginia, P.O. Box 3818, 
Charlottesville, VA 22903-0818}

\begin{abstract} 

We report the discovery and monitoring of radio emission from the Type
Ic \sn{} ranging in frequency from 1.43 to 22.5 GHz, and in time from
4 to 50 days after the SN explosion.  As in most other radio SNe, the
radio spectrum of \sn{} shows evidence for absorption at low frequencies,
usually attributed to synchrotron self-absorption or free-free absorption.
While it is difficult to discriminate between these two processes based on
a goodness-of-fit, the {\it unabsorbed} emission in the free-free model
requires an unreasonably large ejecta energy.  Therefore, on physical grounds 
we favor the synchrotron self-absorption (SSA) model.  In the SSA framework, 
at about day 2,
the shock speed is $\approx 0.3c$, the energy in relativistic electrons
and magnetic fields is $\approx 1.5\times 10^{45}$ erg and the inferred
progenitor mass loss rate is $\approx 5\times 10^{-7}$ M$_\odot$/yr
(assuming a $10^3$ km sec$^{-1}$ wind).  These properties are consistent
with a model in which the outer, high velocity supernova ejecta interact
with the progenitor wind.  The amount of relativistic ejecta in
this model is small, so that the presence of broad lines in the
spectrum of a Type Ib/c supernova, as observed in SN\,2002ap, is not a 
reliable indicator of a connection to relativistic ejecta and hence 
$\gamma$-ray emission.  
\end{abstract}


\section{Introduction}
\label{sec:intro}

Type Ib/c supernovae (SNe) enjoyed a broadening in interest over the
last few years since their compact progenitors (Helium or Carbon stars)
are ideal for detecting the signatures of a central engine.  Such an
engine is expected in the collapsar model (Woosley 1993; MacFadyen,
Woosley \& Heger 2001)\nocite{woo93,mwh01}, the currently popular model
for long-duration $\gamma$-ray bursts (GRBs).  In this model, the
engine (a rotating and accreting black hole) provides the dominant source
of explosive power.  The absence of an extensive Hydrogen envelope in the
progenitor star may allow the jets from the central engine to propagate
to the surface and subsequently power bursts of $\gamma$-rays.

Separately, the Type Ic SN\,1998bw (Galama et al. 1998)\nocite{gvv+98}
found in the localization region of GRB\,980425 (Pian et
al. 2000)\nocite{paa+00} ignited interest in 
``hypernovae''\footnotemark\footnotetext{There is no accepted
definition for a hypernova. Here we use the term to mean a supernova
with an explosion energy significantly larger than $10^{51}$ erg.}.
Regardless of the controversy over the association of SN\,1998bw
with GRB\,980425, or equivalently, the controversy over the relation
of the extremely underluminous GRB\,980425 to the cosmological GRBs,
one fact is not in dispute: {\it SN\,1998bw is a most interesting SN.}
First, the SN exhibited the broadest absorption lines to date, about
60,000 km sec$^{-1}$ (Iwamoto et al. 1998; Woosley, Eastman \&\ Schmidt
1999)\nocite{imn+98,wes99}.  Second, modeling of the optical spectra
and lightcurves suggested a large energy release, $E_{51}\sim 30$,
where $E_{51}$ is the SN energy release in units of $10^{51}$ erg.
Finally, and most relevant to the issue of GRB connection, the SN was the 
brightest radio SN at early times; robust equipartition arguments led to 
an inferred energy
of $E_{\Gamma}\gtrsim 10^{49}$ erg in ejecta with relativistic velocities,
$\Gamma\sim$ few (Kulkarni et al. 1998, hereafter K98).  Until SN\,1998bw,
no other SN showed hints of such an abundance of relativistic ejecta.
Tan, Matzner \&\ McKee (2001)\nocite{tmm01} explain the relativistic ejecta
as resulting from an energetic shock as it speeds up the steep density
gradient of the progenitor. The $\gamma$-ray and radio emission would
then arise in the forward shock.

From the perspective of a GRB--SN connection, what matters most is the
presence of relativistic ejecta. $\gamma$-ray emission traces
ultra-relativistic ejecta, but as was dramatically demonstrated by
SN\,1998bw, the radio serves as an equally good proxy for relativistic
ejecta with the added advantage that the emission is not beamed.
Given this, we began a systematic program of investigating at radio
wavelengths all Ib/c SNe with features similar to SN\,1998bw: a hypernova 
or broad optical lines.

Y. Hirose discovered \sn{} in M74 (distance, $d\sim 7.3\,$Mpc; Smartt
et al.  2002\nocite{}) on 2002, Jan.~29.40 UT (see Nakano
2002)\nocite{n02}.  Mazzali et al. (2002) inferred an explosion date
of 2002, Jan.~$28\pm 0.5$ UT.  Attracted by the broad spectral
features (e.g. Kinugasa et al.  2002; Meikle et
al. 2002)\nocite{kka+02,mls+02} we began observing the SN at
the Very Large Array (VLA\footnotemark\footnotetext{The VLA is
operated by the National Radio Astronomy Observatory, a facility of
the National Science Foundation operated under cooperative agreement
by Associated Universities, Inc.}).

\section{Observations}
\label{sec:obs}

\begin{figure} 
\plotone{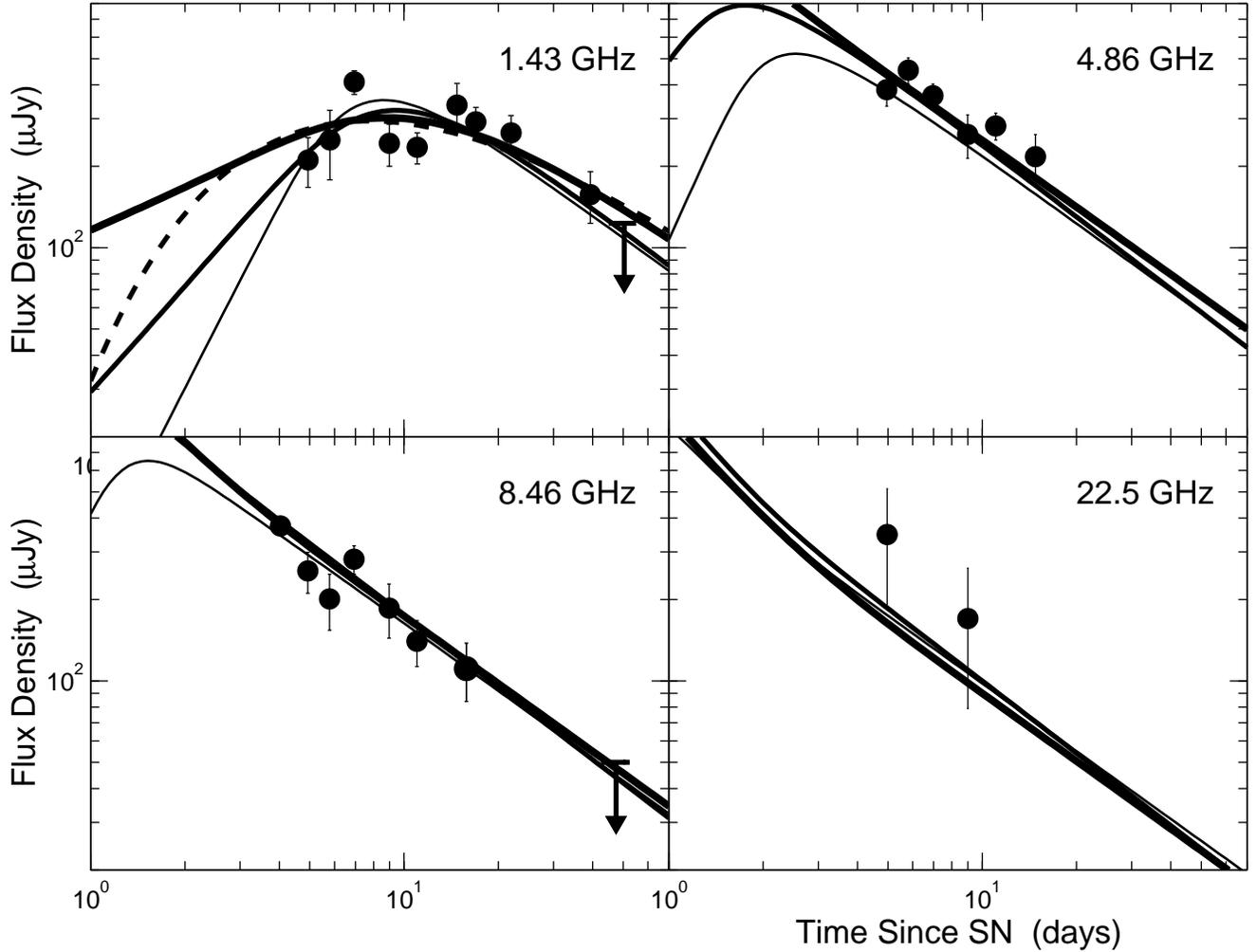}
\caption[]{Radio lightcurves of \sn{}.  The thick solid lines are our
three synchrotron self-absorption models described in \S\ref{sec:SSA},
with $\tau_\nu\propto t^{-1.3}$, $\tau_\nu\propto t^{-2.1}$, and
$\tau_\nu\propto t^{-3}$ in order of decreasing thickness.  The dashed
line is the model-fit based on free-free absorption (\S\ref{sec:FFA}).
At 4.86, 8.46, and 22.5 GHz, the SSA and FFA models provide the same
fit, since the opacity processes do not influence the optically-thin
flux.  The models diverge in the optically-thick regime, which
underlines the importance of rapid, multi-frequency observations.
\label{fig:lcs}}
\end{figure}

We observed \sn{} starting  on 2002, February 1.03 UT.
and detected a radio source coincident with
the optical position at $\alpha$(J2000)= \ra{01}{36}{23.92},
$\delta$(J2000)=\dec{+15}{45}{12.87}, with a 1-$\sigma$ uncertainty of
0.1 arcsec in each coordinate (Berger et al. 2002a)\nocite{bkf02a}.
A log of the monitoring observations and the resulting lightcurves
can be found in Tab.~\ref{tab:obs} and Fig.~\ref{fig:lcs},
respectively.

\subsection{The Radio Spectrum of \sn}
\label{sec:general}

The peak radio luminosity of \sn{}, $L_p(5\,{\rm GHz})
\sim 3\times 10^{25}$ erg sec$^{-1}$ Hz$^{-1}$, is a factor of 20
lower than the typical Ib/c SNe (Weiler et al. 1998)\nocite{wvm+98},
and $\sim 3\times 10^3$ times lower than SN\,1998bw (K98).
The time at which the radio emission peaks
at 5 GHz is $t_p\sim 3$ day, which may be compared with 10 days for
SN\,1998bw (K98)\nocite{kfw+98}, and 10--30 days for the typical Ib/c
SNe (Weiler et al. 1998; Chevalier 1998, hereafter C98)\nocite{wvm+98,che98}.  

The spectral index between 1.43 and 4.86 GHz, $\beta_{1.43}^{4.86}$,
changes from $\sim 0.5$ before day 6, to $\sim -0.3$ at $t\approx 15$
days ($F_\nu\propto \nu^\beta$), while $\beta_{4.86}^{8.46}$ holds
steady at a value of $\approx -0.9$.  This indicates that the spectral
peak, $\nu_p$, is initially located between 1.43 and 4.86 GHz, and
decreases with time.  This peak could be due to synchrotron
self-absorption (SSA) or  (predominantly) free-free absorption (FFA)
arising in the circumstellar medium (CSM).  Regardless of the dominant
source of opacity, the emission for frequencies $\nu>\nu_p$ is from
optically-thin synchrotron emission. 

Massive stars lose matter via strong stellar winds throughout their
life and as a result their CSM is inhomogeneous with density, $\rho(r)
\propto \dot{M}_w v_w^{-1} r^{-2}$. Here, $r$ is the
distance from the star, $\dot{M}$ is the rate of mass loss, and $v_w$
is the wind speed, which is comparable to the escape velocity from the
star.  The progenitors
of Type II SNe are giant stars which have low $v_w\sim 10$ km s$^{-1}$.
Consequently the CSM is dense and this explains why the FFA model has
provided good fits to Type II SNe (e.g. Weiler, Panagia \&\ Montes
2001)\nocite{wpm01}.

On the other hand, the progenitors of Type Ib and Ic SNe are compact
Helium and Carbon stars which have high escape velocities and
therefore fast winds $\sim 10^3\,$km sec$^{-1}$. Thus, {\it a priori},
the CSM density is
not expected to be high.  C98 reviews the modeling of radio lightcurves
of Ib/c SNe and concludes that there is little need to invoke free-free
absorption.  However, synchrotron self absorption is an inescapable
source of opacity and must be included in the modeling of Type Ib/c SNe
(C98; K98)\nocite{che98,kfw+98}.

Low frequency observations provide the simplest way to discriminate
between the two models. In the SSA model, the peak frequency is identified
with the synchrotron-self absorption frequency, $\nu_a$,
and $F_\nu(\nu \ale \nu_a) \propto \nu^{5/2}$. In the FFA model,
the free-free optical depth is unity at the peak frequency, $\nu_{ff}$
and $F_\nu(\nu \ale \nu_{ff})$ decreases exponentially.  Lacking the
requisite discriminatory low frequency data we consider both models.

\subsection{Robust Constraints}
\label{sec:robust}

Before performing a detailed analysis, we derive some general constraints
using the well-established equipartition arguments (Readhead 1994;
K98)\nocite{rea94,kfw+98}.  The energy of a synchrotron source with
flux density, $F_p(\nu_p)$, can be expressed in terms of the
equipartition energy density, $U/U_{\rm eq}=\eta^{11}(1+\eta^{-17})/2$,
where $\eta=\theta_s/\theta_{\rm eq}$, $\theta_{\rm eq}\approx 120d_{\rm
Mpc}^{-1/17}\,F_{\rm p,mJy}^{8/17}\,\nu_{\rm p,GHz}^{(-2\beta-35)/34}$
$\mu$as, and $U_{eq}=1.1\times 10^{56}d_{\rm Mpc}^{2}\,F_{\rm
p,mJy}^{4}\,\nu_{\rm p,GHz}^{-7}\,\theta_{\rm eq,\mu as}^{-6}$ erg.

At about day 7, $F_p(\nu_p=1.4\,{\rm GHz})\approx 0.3$ mJy (see
Fig.~\ref{fig:lcs}).  Thus, $\theta_{\rm eq}(t=7\,{\rm d})\approx 40$
$\mu$as, or  $r\approx 4.5\times 10^{15}$ cm.  The resulting
equipartition energy is $E_{\rm eq}\approx 10^{45}$ erg, the
magnetic field is $B_{\rm eq}\approx 0.2$ G, and the average velocity 
of the ejecta is $v_{\rm eq}\approx 0.3c$ as inferred above.  We note
that any other source of opacity (e.g. free-free absorption) would
serve to increase $\theta_{\rm eq}$, $E_{\rm eq}$, and $v_{\rm eq}$.

\section{A Synchrotron Self-Absorption Model}
\label{sec:SSA}

The synchrotron spectrum from a source with power-law electron
distribution, $N(\gamma)\propto \gamma^{-p}$ for $\gamma>\gamma_{\rm min}$
(\S\ref{sec:general}) is 
\begin{equation}
F_\nu=F_{\nu,0}(\nu/\nu_0)^{5/2}(1-e^{-\tau_\nu})
\frac{F_3(\nu,\nu_m,p)}{F_3(\nu_0,\nu_m,p)}   
\frac{F_2(\nu_0,\nu_m,p)}{F_2(\nu,\nu_m,p)},   
\label{eqn:syn}
\end{equation}
where the optical depth at frequency $\nu$ is given by
\begin{equation}
\tau_\nu=\tau_0(\nu/\nu_0)^{-(2+p/2)}
\frac{F_2(\nu,\nu_m,p)}{F_2(\nu_0,\nu_m,p)},
\label{eqn:tau}
\end{equation}
and
\begin{equation}
F_\ell(\nu,\nu_m,p)=\int_0^{x_m}F(x)x^{(p-\ell)/2}dx;
\label{eqn:F}
\end{equation}
see Li \&\ Chevalier (1999; hereafter LC99)\nocite{lc99}.
Here $x_m\equiv\nu/\nu_m$ (see Rybicki \& Lightman
1979)\nocite{rl79}, and $\nu_m$ is the characteristic
synchrotron frequency of electrons with $\gamma=\gamma_{\rm min}$.
The subscript zero indicates quantities at a reference frequency which
we set to 1 GHz. Finally, $\nu_a$ is defined by the equation
$\tau_{\nu_a}=1$.
 
The evolution of the synchrotron emission depends on a number of
parameters.  Following C98, we assume that $p$, and the fraction of
energy in electrons ($\epsilon_e$) and magnetic fields ($\epsilon_B$)
in the post-shock region remain constant with time; unless otherwise
stated, $\epsilon_e=\epsilon_B=0.1$. The evolution of the synchrotron
spectrum is sensitive to the expansion radius of the forward shock
front, $r_s\propto t^m$, which is related to the density
structure of the shocked ejecta and that of the CSM.  We allow 
for these hydrodynamic uncertainties by 
letting $F_{\nu,0}\propto t_d^{\alpha_F}$ and $\tau_0\propto
t_d^{\alpha_\tau}$, where $t_d$ is the time days since the SN explosion.
In the model adopted here, both these indices depend on $m$ and
$p$. It can be shown that the temporal index of the optically thin flux,
$\alpha=\alpha_F+\alpha_\tau$. The synchrotron characteristic frequency,
$\nu_m$, is particularly useful for inferring the CSM density, and
we parametrize it as $\nu_m=\nu_{m,0}t_d^{\alpha_{\nu_m}}\,$GHz where
$\nu_{m,0}=\nu_m$ GHz.  For typical values of $m$, and $\rho(r)\propto
r^{-2}$, $\alpha_{\nu_m}\approx -0.9$.

With these scalings and Eqs.~\ref{eqn:syn}--\ref{eqn:F} we carry
a least-squares fit to the data. Given the lack of early
optically-thick data (i.e.~1.43 GHz) it is not
surprising that our least-squares analysis allows a broad range of values
for $\alpha_\tau$. In Fig.~\ref{fig:lcs} we plot fits spanning the
minimum $\chi^2$: $\alpha_\tau=-1.3,\,-2.1,\,-3$ (corresponding to
$\chi^2= 40, 43, 46$, respectively and 21 degrees of freedom). 
We note that for other Ib/c SNe $\alpha_\tau$ range from $-2$ to $-3$
(C98; LC99).

\begin{figure} 
\plotone{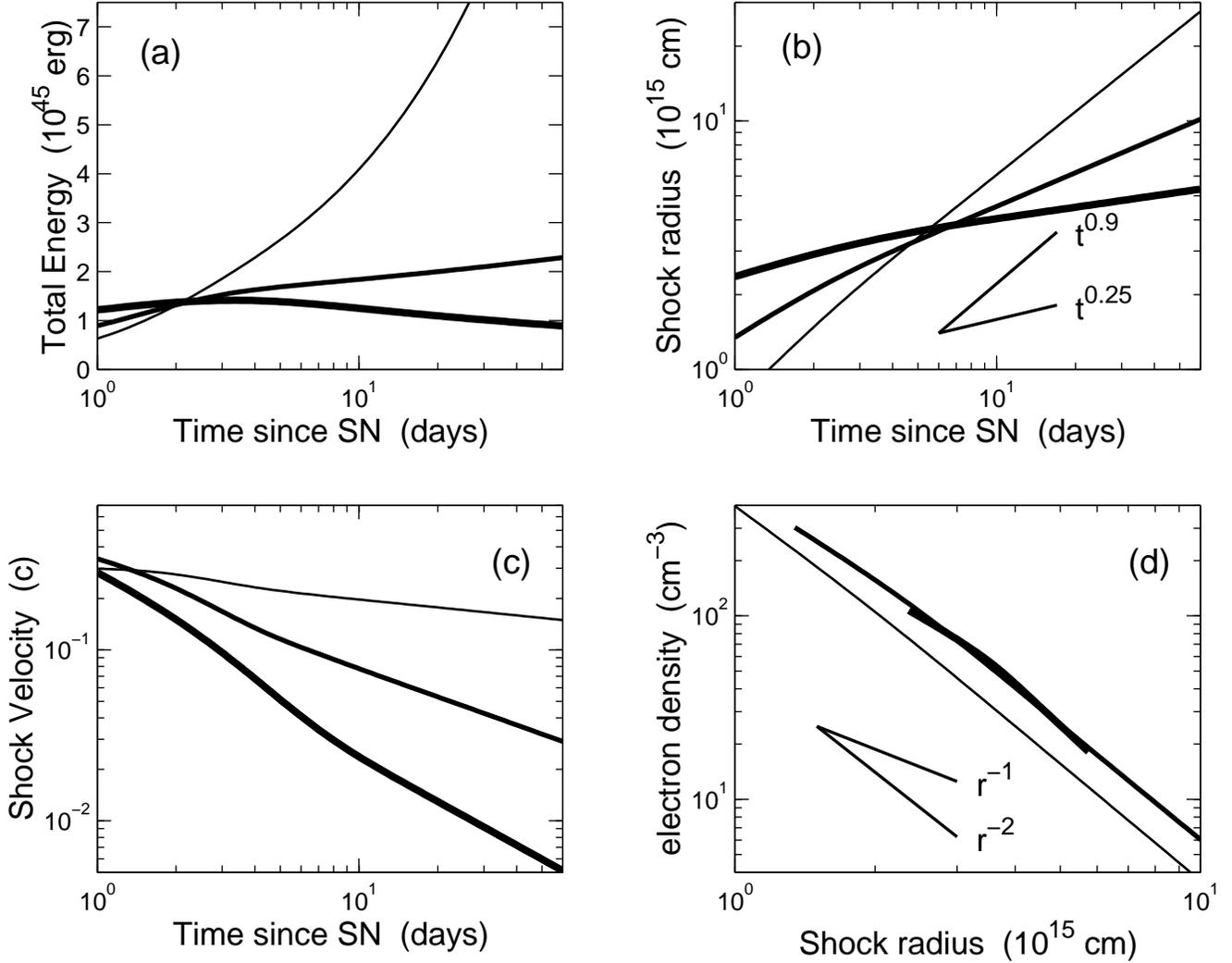}
\caption[]{Inferred physical parameters based on the synchrotron
self-absorption models described in \S\ref{sec:SSA}.  The panels are 
(a) time evolution of the total energy, (b) radius of the radio
photosphere, (c) electron density in the shock as a function of
radius, and (d) velocity of the shock front as a function of time.
Models with $\tau_\nu\propto t^{-1.3}$, $\tau_\nu\propto t^{-2.1}$,
and $\tau_\nu\propto t^{-3}$ are shown in order of decreasing
thickness.  The most likely fit is the one following $r\propto
t^{0.9}$ (i.e.~the model with $\tau_\nu\propto t^{-3}$).
\label{fig:res}}
\end{figure}

The fits, in conjunction with Eqs.~13--15 of LC99 allow us to
trace the evolution of the size of the expanding blastwave
($r_s$), the total (magnetic+electrons) energy ($E$), and the electron
density ($n_e$) in the shock (Fig.~\ref{fig:res}).  We find that for
$\alpha_\tau=-1.3$, $r_s\propto t^{0.25}$ i.e. the blastwave decelerates.
However, $\alpha_\tau=-3$ provides the expected
$r_s\propto t^{0.9}$.  Adopting this physically
reasonable model, we obtain:
$\tau_0(t)=1.2\times 10^3t_d^{-3}$, $F_{\nu,0}(t)=2.9t_d^{2.2}$
$\mu$Jy, and $p=2$.  From Fig.~\ref{fig:res} we note that the early shock
velocity is high, $0.3c$, regardless of the choice of $\alpha_\tau$,
and close to that derived from the simple equipartition arguments
(\S\ref{sec:robust}).

A simple extrapolation of our model, with a cooling break between the
radio and X-ray bands (as expected from the value of $B_{\rm eq}$
inferred in \S\ref{sec:robust}) yields an X-ray flux comparable to
the measurement by XMM-Newton on 2002, Feb 3, $F_X\approx (1.2\pm
0.3)\times 10^{-14}$ erg cm$^{-2}$ sec$^{-1}$ (Rodriguez Pascual et
al. 2002; Soria \& Kong 2002)\nocite{rp+02}.  In addition, the
hardness ratio of the X-ray emission is consistent at the $2\sigma$
level with $p=2$ inferred from the radio data.  
Thus X-ray emission due to other mechanisms (e.g. thermal
brehmsstrahlung) must be quite small.

The mass loss rate of the progenitor star is estimated from $r_s$ and
$n_e$, $\dot{M}_w=8\pi\zeta n_e m_p r^2 v_w\approx 9\times
10^{-9}\nu_{m,0}^{-0.8}$ M$_\odot$ yr$^{-1}$, where the compression
factor is  $\zeta=1/4$,  the nucleon-to-electron ratio is taken to be
2 and $v_w=10^3$ km sec$^{-1}$.  Knowing $B_{eq}$ and our assumed
$\epsilon_e$ we find $\nu_m\sim 10^7\,$Hz and  thus $\dot M_w\approx
5\times 10^{-7}\,M_\odot$ yr$^{-1}$ -- similar to that inferred for
SN~1998bw (LC99).

There are two consistency checks. First, with this $\dot M_w$, 
free-free absorption is negligible.  Second, the kinetic energy of the
swept-up material is $2\times 10^{46}\,$ erg -- consistent with our
estimate of the equipartition energy and $\epsilon_e$.

\subsection{The SSA Model in the Context of a Hydrodynamic Model}
\label{sec:hydro}

The results of \S\ref{sec:SSA} can be tied in to a fairly simple
hydrodynamic model.  Matzner \& McKee (1999)\nocite{mm99} show that
for the 
progenitors of Ib/c SNe (compact stars with radiative envelopes) the ejecta
post-explosion density profile can be described by power laws at low
and high velocities, separated by a break velocity, $v_{\rm ej,b}=5150
(E_{51}/M_1)^{1/2}\approx 2\times 10^4$ km sec$^{-1}$; where
the mass of the ejecta is $M_{\rm ej}=10M_1\,M_\odot$.  We use $E_{51}\approx 
4-10$ and $M_1\approx
0.25-0.5$ for \sn{} (Mazzali et al. 2002).  
At $v_s\approx 0.3c$, the density profile is given by $\rho_s\approx
3\times 10^{96}E_{51}^{3.59}M_1^{-2.59}t^{-3} v^{-10.18}$ g
cm$^{-3}$.  This profile extends until radiative losses become important
when the shock front breaks out of the star. Using Eq. 32 of Matzner \&\
McKee (1999)\nocite{mm99} this happens for $v_s\approx 1.5c$ (assuming
a typical 1 R$_\odot$ radius for the progenitor star).  Thus, the
outflow can become relativistic.

Using the self-similar solution of Chevalier (1982)\nocite{c82} 
the velocity of the outer shock radius, $R$, (assuming a $\rho=A
r^{-2}$ CSM) is 
\begin{equation}
{R\over t}=52,300\, E_{51}^{0.44} M_1^{-0.32} A_*^{-0.12} t_d^{-0.12}
{\rm~km~sec^{-1}},
\end{equation}
where $A_*=(\dot M_w/10^{-5}~M_\odot {\rm~yr^{-1}})
(v_w/10^3{\rm~km~sec^{-1}})^{-1}$. The shock velocity, $\dot R$ is
insensitive to the circumstellar wind density. Thus, we find that the
velocities inferred from the radio observations of \sn{} can be
naturally accounted for by the outer supernova ejecta.

The energy above some velocity $V$ is \begin{equation} E(v>V)\approx
\int_V^{\infty}{1\over 2}\rho_fv^2 4\pi v^2t^3dv =7.2\times
10^{44}E_{51}^{3.59}M_1^{-2.59}V_5^{-5.18} {\rm~ergs}, \end{equation}
where $v_5$ is the velocity in units of $10^5 {\rm~km~s^{-1}}$.
For the preferred SN 2002ap parameters, $E(v>V)\approx 4.2\times
10^{48}V_5^{-5.18}$ erg.  There is therefore plenty of energy in the
high velocity ejecta to account for the observed radio emission.

Indeed, given the over-abundance of $E(v>V_5)$ relative to the energy
inferred from the radio emission, we wonder how secure the are the estimates of
$E_{51}$ and $M_1$ of Mazzali et al. (2002). In particular, $E_{51}$ and $M_1$ 
are derived from early time optical observations and are certainly subject to
asymmetries in the 
explosion.  For SN~1998bw, the asymmetric model of H\"oflich, Wheeler \&\
Wang (1999)\nocite{hww99} yielded $E_{51}\sim 2$, an order of magnitude
smaller than that obtained from symmetrical models (e.g. Iwamoto et
al. 1998\nocite{imn+98}).

\subsection{Interstellar Scattering \&\ Scintillation}
\label{sec:iss}

Interstellar scattering and scintillation (ISS) is expected
for radio SNe (see K98). Indeed, the perceptible random deviations from
the model curves (see Fig.~\ref{fig:lcs}), which account for the high
$\chi_{\rm min}^2$ could arise from ISS.

We compute the ISS variability, specifically the modulation index (the
ratio of the rms to the mean) of the observed flux, using the ISS model
of Goodman (1997)\nocite{goo97}.  Along the direction to \sn\, using the
Galactic free electron model of Taylor \& Cordes (1993)\nocite{tc93} we
find a scattering measure, ${\rm SM}_{-3.5}\approx 0.65$ m$^{-20/3}$ kpc
and the distance to the scattering screen of $d_{\rm scr}\approx 0.5$ kpc.

For radio SNe there are two basic regimes of interest separated by the
so-called transition frequency, $\nu_{\rm tr}$.  For
$\nu>\nu_{\rm tr}$, the {\it weak scattering} regime, the
modulation index is $m_{\rm \nu,w}=(\nu_{\rm tr}/\nu)^{17/12}$ [for
$\theta_s=r_s/d$ smaller than the Fresnel size,
$\theta_{\rm F}=8.1(d_{\rm scr}\nu_{\rm GHz})^{-1/2}$ $\mu$as] and 
$m_{\rm \nu,q}=m_{\rm \nu,w}(\theta_{\rm F}/\theta_{\rm s})^{7/6}$
otherwise.  For $\nu<\nu_{\rm tr}$, the {\it strong (refractive)
scattering} regime, $m_{\rm
\nu,R}=(\nu/\nu_{\rm tr})^{17/30}$ [for $\theta_s<\theta_{\rm
R}=\theta_{\rm F,tr}(\nu_{\rm tr}/\nu)^{11/5}$, where $\theta_{\rm
F,tr}$ is the Fresnel size at $\nu_{\rm tr}$], and $m_{\rm
\nu,q}=m_{\rm \nu,R}(\theta_{\rm R}/\theta_{\rm s})^{7/6}$ otherwise.

For \sn\ we find $\nu_{\rm tr}\approx 7.3$ GHz.  From
Fig.~\ref{fig:res} the source angular diameter, over the period 5--20
days, is $\theta_s\approx 60-160$ $\mu$as, from which we estimate the 
following  values: $m_{8.46}\approx 5\%$, $m_{4.86}\approx
10\%$, and $m_{1.43}\approx 40\%$.

We estimate the actual modulation index empirically by adding $m_\nu
F_\nu$ in quadrature to each measurement error so that the reduced
$\chi_{\rm min}^2$ is unity.  Here $F_\nu$ is the model flux described
in \S\ref{sec:SSA}.  We find $m_{8.46}\approx 0.1$, $m_{4.86}\approx
0.2$, and $m_{1.43}\approx 0.3$, in good agreement with the
theoretical estimates.  This provides an independent confirmation of
the size, and hence expansion velocity of the ejecta.

\section{A Free-Free Absorption Model}
\label{sec:FFA}

In this model, the spectrum is parametrized as
(Chevalier 1984; Weiler et al.
1986)\nocite{che84,wsp+86}:
\begin{eqnarray} 
F_\nu=K_1\nu_5^\alpha t_d^\beta\,e^{-\tau_\nu} \nonumber \\
\tau_\nu=K_2\nu_5^{-2.1} t_d^\delta, 
\label{eq:FFA}
\end{eqnarray}
where $\nu_5=5\nu$ GHz.  We find an acceptable fit
($\chi^2=40$ for $21$ degrees of freedom) yielding:
$K_1\approx 2$ mJy, $K_2\approx 0.4$,
$\alpha\approx -0.9$, $\beta\approx -0.9$, and $\delta\approx -0.8$.
With these parameters and Eq.~16 of Weiler et al. 1986\nocite{wsp+86}) we
find $\dot{M}_w\approx 5\times
10^{-5}$ M$_\odot$ yr$^{-1}$ for $v_w=10^3$ km sec$^{-1}$.

In this model, one day after the explosion $\nu_{\rm 
ff}\approx 3.2$ GHz, and $F_\nu(\nu_{\rm
ff})\approx 1.1$ mJy (Fig.~\ref{fig:lcs}).  
The unabsorbed flux at the peak of the synchrotron spectrum is 
$F_\nu(\nu_a)\approx 3(\nu_a/3.2\,{\rm
GHz})^{-0.9}$ mJy (note $\nu_a<\nu_{ff}$ in the FFA model) 
for which
$r_{\rm eq}\approx 7.5\times 10^{15}(\nu_a/3.2\,{\rm
GHz})^{-3/2}$ cm. Thus  $v_{\rm eq}\approx 3c(\nu_a/3.2\,{\rm
GHz})^{-3/2}$, which corresponds to $\Gamma=2(\nu_a/3.2\,{\rm
GHz})^{-1}$ if relativistic effects are taken into account (R. Sari
priv.~comm.).  Alternatively, if we fix the expansion velocity to the 
optical value, $v_s\approx 3\times 10^4$ km sec$^{-1}$ (Mazzali et
al. 2002), we find a brightness temperature, $T_b\approx 4\times
10^{13}$ K --- clearly in excess of the equipartition temperature,
again necessitating a high bulk Lorentz factor, $\Gamma\sim 10^2$.  

Thus, even if $\nu_a=\nu_{\rm ff}$ (in which case  free-free opacity
would not be necessary in the first place), the FFA model requires truly
relativistic ejecta, or alternatively a large departure from
equipartition, resulting in $E\approx 7\times 10^{50}(\nu_a/3.2\,{\rm
GHz})^{-9}$ erg (for $v_s\approx 0.5c$ instead of $3c$).  Clearly,
the energy requirement would increase by many orders of magnitude if
$\nu_a\ll \nu_{\rm ff}$.

\section{Discussion and Conclusions}
\label{sec:conc}

The type Ic SN\,1998bw, likely associated with GRB\,980425,
was peculiar in two ways. It  exhibited broad photospheric absorption 
lines, and exceedingly strong radio emission at early times.  The latter
was interpreted to arise from relativistic ($\Gamma\sim$ few) ejecta of 
total energy $E_\Gamma\age 10^{49}$ erg.  
These two peculiarities made 
sense in that the simple theory suggested that broad photospheric features 
are a reliable indicator of relativistic ejecta, a necessary condition for 
$\gamma$-ray emission.

The type Ic SN\,2002ap elicited much interest because it too displayed
similar broad lines.  However, from our radio observations we estimate the 
energy in relativistic electrons and magnetic fields to be quite modest: 
$E\approx 1.5\times 10^{45}$ ergs in ejecta with a velocity $\approx 0.3c$.  
In view of this, the absence of $\gamma$-rays from \sn{} is not surprising
(Hurley et al. 2002)\nocite{hur02}.  Both the energy and speed of the 
ejecta can be accounted for in the standard hydrodynamical model.  We thus 
conclude that broad photospheric lines are not good predictors of relativistic 
ejecta.

Separately, the broad photospheric features led modelers to conclude that 
\sn{} was a hypernova with an explosion energy of $E_{51}\sim 4-10$ erg
(Mazzali et al. 2002).  
However, the radio observations paint a different picture.  The low $E_\Gamma$ 
make \sn{} an ordinary Type Ib/c SN (or perhaps even a low energy event; 
\S\ref{sec:general}).  We also draw attention to the significant role played 
by asymmetries in forming the optical lines which, when not properly 
accounted for can lead to discrepant estimates (cf.~the large variations
in the estimated SN energy release for SN\,1998bw; see H\"oflich, Wheeler 
\& Wang 1999\nocite{hww99}; Iwamoto et al. 1998\nocite{imn+98}).  Thus, we 
wonder how reliable are the inferred $E_{51}$ and $M_1$ values.  Along the 
same vein, we note that Kawabata et al. (2002)\nocite{kji+02} suggest, based 
on spectro-polarimetric observations, a jet with a speed of $0.23c$ and 
carrying $2\times 10^{51}$ erg.  Such a jet, regardless of geometry, would 
have produced copious radio emission. 
 
We end with three conclusions.  First, at least from the perspective of 
relativisic ejecta, \sn{} was an ordinary Ib/c SN.  Second, broad photospheric 
lines appear not to be a good proxy for either an hypernova origin or 
$\gamma$-ray emission.  Third, radio observations offer a practical and 
accurate proxy for relativistic ejecta.

\acknowledgments
Dale Frail was involved in various aspects of this project and we are grateful
for his help and encouragement.  We also wish to acknowledge useful discussions
with J. Craig Wheeler.  Finally, we thank NSF and NASA for supporting our research.

\clearpage
\begin{deluxetable}{lllll}
\tabcolsep0.1in\footnotesize
\tablewidth{\hsize}
\tablecaption{Radio Observations of \sn{} \label{tab:obs}}
\tablehead {
\colhead {Epoch}      &
\colhead {$F_{1.43}\pm\sigma$} &
\colhead {$F_{4.86}\pm\sigma$} &
\colhead {$F_{8.46}\pm\sigma$} &
\colhead {$F_{22.5}\pm\sigma$} \\
\colhead {(UT)}      &
\colhead {($\mu$Jy)} &
\colhead {($\mu$Jy)} &
\colhead {($\mu$Jy)} &
\colhead {($\mu$Jy)} 
}
\startdata
2002 Feb 1.03 & \nodata & \nodata & $374\pm 29$ & \nodata \\
2002 Feb 1.93 & $211\pm 44$ & $384\pm 50$ & $255\pm 44$ & $348\pm 165$\\
2002 Feb 2.79 & $250\pm 72$ & $453\pm 50$ & $201\pm 47$ & \nodata \\
2002 Feb 3.93 & $410\pm 41$ & $365\pm 38$ & $282\pm 34$ & \nodata \\
2002 Feb 5.96 & $243\pm 43$ & $262\pm 48$ & $186\pm 42$ & $170\pm 91$ \\
2002 Feb 8.00 & $235\pm 31$ & $282\pm 32$ & $140\pm 27$ & \nodata \\
2002 Feb 11.76 & $337\pm 68$ & $217\pm 45$ & $111\pm 27$ & \nodata \\
2002 Feb 13.94 & $292\pm 38$ & \nodata & \nodata & \nodata \\
2002 Feb 18.95 & $266\pm 42$ & \nodata & \nodata & \nodata \\
2002 Mar 4.85+11.83 & $157\pm 34$ & \nodata & \nodata & \nodata \\
2002 Mar 18.77+19.97 & $57\pm 33$ & \nodata & $25\pm 25$ & \nodata 
\enddata
\tablecomments{The columns are (left to right), (1) UT date of each
observation, and flux density and rms noise at (2) 1.43 GHz, (3)
4.86 GHz, (4) 8.46 GHz, and (5) 22.5 GHz.  Observations with more than
one date have been co-added to increase the signal-to-noise of the
detection.}   
\end{deluxetable}

\clearpage

\clearpage

\end{document}